\documentclass[a4paper,12pt,reqno,superscriptaddress,nofootinbib]{revtex4}
\usepackage[centertags]{amsmath}
\usepackage{amsfonts}
\usepackage{amssymb}
\usepackage{amsthm}
\usepackage{newlfont}
\usepackage{stmaryrd}
\usepackage{mathrsfs}
\usepackage{euscript}
\usepackage{siunitx}
\usepackage{physics}
\usepackage{algorithm2e}
\usepackage{graphicx,subcaption}
\usepackage{enumitem}
\usepackage{natbib}
\usepackage{graphicx}
\usepackage{color}
\usepackage{floatrow}
\usepackage{caption}
\usepackage{hyperref}
\usepackage{hhline}
\usepackage{hyperref}
\usepackage{cleveref}
\usepackage{import}
\theoremstyle{plain}
  \newtheorem{theorem}{Theorem}[section]

\theoremstyle{definition}

\theoremstyle{remark}

  \newtheorem{example}[theorem]{Example}

\numberwithin{equation}{section}

\newcommand{\caF}{{\mathcal F}}

\newcommand{\bbR}{{\mathbb R}}

\newcommand{\opunit}{\text{1}\kern-0.22em\text{l}}

\DeclareMathAlphabet{\mathpzc}{OT1}{pzc}{m}{it}

\newcommand{\id}{\textrm{d}}

\usepackage{floatrow}
\usepackage{caption}
\usepackage{color,soul}
\DeclareCaptionJustification{justified}{\leftskip=0pt \rightskip=0pt \parfillskip=0pt plus 1fil}
\begin{document}

\title{Relaxation to nonequilibrium}

\author{Christian Maes}\affiliation{Department of Physics and Astronomy, KU Leuven, Belgium.}
\author{Karel Neto\v{c}n\'{y}}
\affiliation{Institute of Physics, Czech Academy of Sciences, Prague, Czech Republic}
\keywords{macroscopic fluctuation theory, nonequilibrium response, relaxation to nonequilibrium, extending gradient flow and {\tt GENERIC}.}

\begin{abstract}
We describe the structure of relaxation for a steadily driven macroscopic body. The time-evolution is characterized as the zero-cost flow for a nonequilibrium and nonlinear extension of the Onsager-Machlup action governing the dynamical fluctuations.
The approach hinges on two main elements: the principle of local detailed balance, which identifies the relevant thermodynamic forces, and the canonical decomposition of the frenesy into a Legendre pair. Notably, it is the time-symmetric component of the Lagrangian, the frenesy, that shapes the structure of the macroscopic evolution for given forcing. We add a simple argument for why the nonequilibrium entropy, which governs the {\it static} macroscopic fluctuations  of the system, is monotone in time.
The results can be interpreted as the steady nonequilibrium extension of {\tt GENERIC} where relaxation to equilibrium is governed by a dissipative gradient flow superimposed on a Hamiltonian flow. 
\end{abstract}

\maketitle

%\tableofcontents  %to delete when submitting

\section{Introduction}
For macroscopic systems, the question of relaxation to equilibrium stands at the beginning of nonequilibrium physics.  The problem is obviously foundational for the dynamical characterization of equilibrium. Consistency with the First and Second Laws of thermodynamics makes important constraints.  Those are realized in the structure of {\tt GENERIC}, acronym of {\it General Equation for Non-Equilibrium Reversible-Irreversible Coupling}, which covers a very wide range of examples, \cite{complexfluids1,complexfluids2,genericmaes}.  A subclass contains the gradient flow dynamics, where the motion chooses steepest ascent/descent, and a (convex or concave) cost function (free energy or entropy) gets extremized.  For instance, the homogeneous Boltzmann equation is a gradient flow, the Vlasov–Fokker–Planck equation for underdamped diffusions is {\tt GENERIC}, {\it etc}.  On the whole, in applied mathematics, variational evolution equations are a widely used and efficient tool to target local minima of a cost function or a thermodynamic potential.\\

Much less understood (or even studied) is the structure and characterization of macroscopic evolutions toward a stationary {\it non}equilibrium condition, \cite{Maes2026WhatIsNonequilibrium}.
In general, these are nonvariational. We have in mind the relaxation behavior of open macroscopic systems that are subject to rotational forces, or of chemical reactors where there are input- and output-flows. The relaxation to nonequilibrium can be much richer than to equilibrium, including the convergence to limit cycles and turbulent or chaotic behavior.\\ 
It is important to distinguish that program from the context of (traditional) irreversible thermodynamics that assumes local equilibrium, \cite{dGM,gasp,Spohn1991,prig,GP}.  
%{\it E.g.}, we do not consider here diffusive dynamics where the source of nonequilibrium is found in conflicting boundary conditions, {\it e.g.} different temperatures or densities at the edges of a volume.  Neither do we follow irreversible thermodynamics where ``equilibrium'' evolutions for microscopically large but macroscopically small domains are used under the assumption of local equilibrium, \cite{Spohn1991,prig,GP}.\\
Instead, the framework of the present paper is a direct adaptation to nonequilibrium situations of the fluctuation theory for {\tt GENERIC}; see \cite{Kraaij_2020,genericmaes}.  
We consider a formulation entirely in terms of thermodynamic forces (in the sense of Onsager), also for describing the influence of boundary conditions, where driving is inserted via the condition of local detailed balance.  
From a different perspective, it proposes a general dynamical macroscopic fluctuation theory (incorporating the one for diffusive systems as a special case, \cite{M,F,T}).
Indeed, we connect the structure of dynamical fluctuations with the structure of relaxation, much in the spirit of Onsager theory \cite{pons,O,MPR13,Kirkwood1946}.\\

In Section \ref{men} we present the main observation, avoiding mathematical sophistication and postponing derivations.  In particular, it is explained how the result fits the Onsager perspective combining the relaxation dynamics with dynamical fluctuations. We discuss the monotonicity of the nonequilibrium entropy and how the results contain results for boundary driven diffusion processes of interacting particle systems. The more detailed arguments make Section \ref{set}.   In Section \ref{cans} we exploit the canonical structure of the action, as expressed before in \cite{sdiv,maes2007entropy,Maes_2008}, to derive the canonical form of return to stationary macroscopic nonequilibrium.  The relaxation equation appears in Section \ref{rele}. That is followed by a short discussion in Section \ref{const} on constructions of macroscopic relaxation toward nonequilibrium, independent of detailed microscopic modeling.  
We end with illustrations in Section \ref{exam}, making the formul{\ae} concrete for some physical examples.

\section{Main result}\label{men}
%The present section gives a summary of the main result.  Following sections are mostly mathematical elaborations and physics examples.  The present section and its discussion suffice in the first reading.\\

\subsection{Structure of dynamical fluctuations}
An autonomous equation for the state $z$ of a physical system is obtained by specifying how the time-derivative or displacement $\dot z$ depends on $z$. The latter may refer to spatial profiles or to the densities of various species in the limit where the number of particles $N$ tends to infinity.\\
Before taking the macroscopic limit, there are in general many possible paths $(z(s),j(s)), s\in [0,t],$ of time-dependent states and currents related via $\dot z(s) = D j(s)$ for some operator $D$ (such as minus the divergence).  Which paths are more plausible depends on the  kinetics, interactions, and driving. All that gets summarized in the Lagrangian $\cal L(j,z)\geq 0$ that governs the probabilities of dynamical fluctuations in the sense that
\begin{equation}\label{prob}
\text{Prob}\big[(z(s),j(s)), s\in [0,t]\big] \sim e^{-N\,\int_0^t \id s \,\cal L(j(s),z(s))\,}
\end{equation}
renders the conditional probability of possible paths, ignoring for now the initial macroscopic statistics. Note that we call $\cal L$ a Lagrangian, but so far without ``mechanical'' motivation.  To obtain an explicit \eqref{prob} from more microscopic considerations requires understanding the fluctuations around a law of large numbers, which we ignore here.  Instead, we take the road (in the form of path-space considerations) as pioneered in the Onsager-Machlup theory, \cite{Onsager1951,Onsager1953}, but generalized to nonequilibrium nonlinear evolutions.\\
In the macroscopic limit $N\uparrow \infty $, the ``true'' or deterministic updating rule {\it emerges} in the form $\dot z = D j_z$, and $j_z$ is found by minimizing $\cal L(j,z)$ over $j$ (zero-cost limit), that is, maximizing probability \eqref{prob}.  Once we know the structure of the Lagrangian in \eqref{prob}, we will find  $j_z$, {\it i.e.}, determining the macroscopic equation $\dot z = D j_z$.  That is similar but not identical to characterizing motion from the Principle of Least Action; see for instance \cite{aaron} for a comparison.\\

To derive the form of the Lagrangian and hence the structure of its zero-cost limit, we assume local detailed balance \cite{time,ldb}.  That means we suppose there is a ``Hamiltonian'' flow $J^H$ and ``force'' $F$, functions of the state $z$ with $J^H(z)\cdot F(z) = 0$, so that for all $(j,z)$,
\begin{equation}\label{lbd}
\cal L\left(J^H(z)- j, z \right) - \cal L\left(J^H(z) + j, z \right) = j\cdot F(z)
\end{equation}
In other words, the antisymmetric contribution to the Lagrangian is given by the entropy production (current times force), as first pronounced in \cite{maes1999fluctuation,maes2000entropy} for $J^H=0$.  The nonequilibrium aspect is in the fact that the force $F$ does not need to be the gradient/derivative of a thermodynamic potential.  A typical situation is that $F$ sums over multiple thermodynamic forces arising from coupling the system with different equilibrium baths (in terms of temperature or chemical potential and/or rotational forces).  We will not always indicate the functional dependence of the Lagrangian via $\cal L = 
\cal L_F$ on the force $F$, but it is of course important to remember.\\
Perhaps surprising, we will see in Section \ref{cans} that under the sole condition of local detailed balance \eqref{lbd}, the Lagrangian $\cal L$ in \eqref{prob} can always be written as 
\begin{eqnarray}
\cal L(j,z) &=& \psi(j-J^H(z),z)+\psi^\star(\frac{F(z)}{2},z)-(j-J^H(z))\cdot \frac{F(z)}{2}\nonumber\\
\cal L(j+ J^H(z),z) &=& \psi(j,z)+\psi^\star(\frac{F(z)}{2},z)-j\cdot \frac{F(z)}{2}\label{LGEN}
\end{eqnarray}
where $(\psi,\psi^*)$ are a Legendre pair (which typically can depend on $F$).\\

Various aspects of the canonical structure for nonequilibrium dynamical fluctuations above have already been observed in \cite{Maes_2008}.
We emphasize that this structure of the functional of dynamical fluctuations is completely general, and has various realizations, including the macroscopic fluctuation theory for either jump or diffusion processes, \cite{M,F,T,Maes_2008,Maes2008Steady,maes2007entropy}. We also mention the more recent \cite{PhysRevResearch.4.033208} for the interesting case of nonequilibrium  chemical reaction networks.  (We thank Roman Belousov for pointing out this reference, after our work was completed.)\\
  We refer to Lemma 2.1 in \cite{MPR13} for the case $J_H=0$ and in the context of reversible Markov processes.  In \cite{Kaiser2018}, \eqref{lbd}--\eqref{LGEN} was studied without $J^H$, but with nongradient $F$, following up on the canonical decomposition in \cite{Maes_2008,maes2007entropy,Maes2008Steady}. Similarly, the local detailed balance condition \eqref{lbd} (with $J^H=0$) has appeared in Chapter 10.5.3 of  \cite{Hoeksema2023}.

\subsection{Structure of relaxation equation}

From minimizing \eqref{LGEN}, {\it i.e.}, putting $\cal L(j,z) = 0$, as a simple consequence of Legendre duality (see Section \ref{rele}), the relaxation equation must have the form
\begin{equation}\label{rel}
\dot z=D\,J^H(z) + D\,\partial_f\psi^\star(\frac{F(z)}{2},z)
\end{equation}
where $\partial_f\psi^\star$ is the derivative with respect to the first argument in the convex function $f\mapsto \psi^*(f,z)$.  The notation is illustrated in the Examples of Section \ref{exam}.  The structure \eqref{rel} unifies in a physical way a wide variety of equations that describe relaxation to macroscopic nonequilibrium.\\
We repeat that the Lagrangian and hence $\psi^*$ can also depend on the force $F$, in particular on its rotational part (see example \eqref{cho} below).  In other words, a functional dependence of $\psi^*(f,j) = \psi^*_F(f,j)$  makes a double dependence in \eqref{rel} of the current on the driving.
%\begin{equation}\label{relf}
%\dot z=D\,J^H(z) + D\,%\partial\psi^\star_F(\frac{F(z)}{2},z)
%\end{equation}
 Note in particular that the double dependence on the nonequilibrium force $F$ is responsible for the violation of Onsager-reciprocity in the current {\it vs} force characteristic around nonzero nonequilibrium forcing.

\subsection{Nonequilibrium entropy}
 
  When the thermodynamic force $F(z) = D^\dagger \id S(z)$ in \eqref{rel} can be derived from a (free) entropy $S$, where $D^\dagger$ is the adjoint of $D$ and $\id S$ is the gradient in the $z-$space, we recover the relaxation to a stable equilibrium as for {\tt GENERIC}.   Indeed, relaxation to equilibrium is the special case where in \eqref{rel}, $F =D^\dagger \id S$ and $D\,J^H \cdot \id S=0$. Then,
\begin{equation}
\dot S = \id S\cdot \dot z = 2D^\dagger\id S/2\cdot\partial_f \psi^*(D^\dagger \id S/2) \geq 0
\end{equation}
where the last inequality uses $\psi^*(f,z) \geq \psi^*(0,z) =0$.  In other words, $S$ never decreases in time.\\
The dynamical system \eqref{rel} extends the framework of {\tt GENERIC} to relaxation toward nonequilibrium steady conditions.
Interestingly, the monotonicity remains true for steady nonequilibrium, where it is the nonequilibrium entropy $S$ (per $k_B$) that never increases, even though the thermodynamic force $F$ is not derived from a potential.  More specifically, that nonequilibrium entropy governs the stationary static fluctuations, much in the spirit of the Boltzmann-Einstein formula for equilibrium systems,
\begin{equation}\label{invr}
\text{Prob}[z] \sim e^{N S(z)}
\end{equation}
where ``Prob'' refers to the stationary distribution, {\it i.e.}, the invariant law for \eqref{prob}. For an environment at one fixed temperature where the driving is adding a rotational force, the $S$ can be called a nonequilibrium Massieu function.  In thermal equilibrium, for one particle bath at chemical potential $\mu$ and for $z =$ the density profile of a gas, we would have
$S(z) N = \beta \,V \,[\Omega(\mu) - \Omega(\mu,z)]$, where $\Omega(\mu,z)= \caF(z) - \mu z$, with $\Omega$ the grand potential, and $\caF$ the free energy functional.\\ 
We present the monotonicity argument for $J^H=0$.
Following the structure of Freidlin-Wentzel theory \cite{fw,DemboZeitouni1998}, by the invariance of \eqref{invr} reached asymptotically in time, this nonequilibrium entropy (aka quasipotential) $S$ is a solution of the Hamilton-Jacobi equation
\begin{equation}
\cal H(-D^\dagger \frac{\id S}{\id z},z) =0
\end{equation}
where $\cal H$ is the Hamiltonian and Legendre dual of the Lagrangian
$\cal L$.  In other words,
\begin{equation}
\sup_j\{ -j\cdot D^\dagger \frac{\id S}{\id z} - \cal L(j,z)\} =0
\end{equation}
Hence, in particular, the zero-cost flow $j_z$ satisfies
\[
-j_z\cdot D^\dagger \frac{\id S}{\id z} - \cal L(j_z,z) =  -j_z\cdot D^\dagger \frac{\id S}{\id z}\leq 0
\]
and we conclude that 
\begin{equation}
\dot S = \frac{\id S}{\id z}\cdot \dot z = D^\dagger\frac{\id S}{\id z}\cdot j_z \geq 0
\end{equation}
giving the dynamical monotonicity of this nonequilibrium entropy (which, we remind, is a state function).
Another argument can be taken from Section 7 in \cite{GP}; see also \cite{BodineauLebowitzMouhotVillani2013,DeRoeckMaesNetocny2006,T}.  Note that we speak here about a single macroscopic body and not about the average behavior in an ensemble of identical and independent copies of a (possibly small) system (where $S$ is a relative entropy as follows from Sanov's theorem).\\
There is an operational meaning to this nonequilibrium entropy when close enough to equilibrium; see e.g. \cite{Bertini_2013,BasuMaesNetocny2015}.  The reason is the form of the McLennan distribution, which is valid there, in terms of excess work. 

\subsection{Remarks}

{\bf 1:} Note that apart from the choice of macroscopic variable (allowing an autonomous dynamics), the input for \eqref{rel} is threefold: we need (1) to identify the Hamiltonian flow $J^H$ and the thermodynamic force $F$, (2) we need to understand the operator $D$, mostly just minus the divergence, the unit operator or some stocheometric matrix, and (3) we need the convex function $\psi^*$.  {\it Ad} (1), $J^H$ and $F$ follow from local detailed balance.  {\it Ad} (2), the operator $D$ gets specified from the kinematic relation between the temporal change in the state, {\it aka} the displacement, and the current. Finally,  {\it ad} (3), $\psi^* = \psi^*_F$ is determined by the frenetic part in the Lagrangian, \cite{fren}.
 In particular, the universal structure \eqref{rel} highlights the role and importance of the time-symmetric part in the Lagrangian, and how it gets influenced by the forcing $F$. Indeed, complementary to \eqref{lbd}, we have the time-symmetric contribution
\begin{equation}\label{freco}
\frac 1{2}\left[\cal L\left(J^H(z)- j, z \right) + \cal L\left(J^H(z) + j, z \right) \right]= \psi(j,z) + \frac 1{2}\psi^\star(\frac{F(z)}{2},z)
\end{equation}
where $\psi(j,z) = \psi(-j,z)$ is the Legendre transform of the function $\psi^*(f,z)$.\\
%The second main result is obtained by exploiting the fact that the relaxation \eqref{relf} carries information about the path-space probability and {\it vice versa}: from the above, the $\psi^*$ basically determines both $\cal L$ and the relaxation equation.  It allows an extension of the Onsager-Machlup arguments.  Details will be given in Section \ref{flu}.\\

{\bf 2:}  When, from observing relaxation, one identifies the function $\psi^*_F$ and the force $F$, one can reconstruct the Lagrangian and hence the dynamical fluctuations.  {\it Vice versa}, when one knows \eqref{freco}, the frenetic contribution to the fluctuations, one gets the function $\psi^*_F$, and hence the relaxational behavior.  That is the strongest instance of a fluctuation-response correspondence. In other words, moving between \eqref{prob} and \eqref{rel}, one connects fluctuations and response.  That will be exposed in more detail for response relations in an upcoming paper, which extends the Onsager theory (1931) to nonlinear and nonequilibrium dynamics, \cite{pons,O,Kirkwood1946}.\\

{\bf 3:} The fluctuation structure \eqref{LGEN} extends the one known before from {\it macroscopic fluctuation theory}, \cite{M,F,T}.  In the case of boundary driven diffusions, we have the structure
   \begin{eqnarray}
{\cal L} (z,j) &=& \frac 1{4} \int \id r [j- j_z]\cdot\chi(z)^{-1}[j- j_z]\notag\\
&=& \frac 1{4} \int \id r \frac{|j+{\cal D}(\rho) \nabla \rho|^2}{\chi(\rho)}\notag
\end{eqnarray}
Here,
$j_z $ is the actual zero-cost flow (the hydrodynamic current), with $\dot z = Dj$ which means $\partial_t \rho + \text{ div }j =0$, when
 $z=\rho$ is macroscopic density profile of particles and $j$ the associated current. We write
 ${\cal D}(\rho) =$ for the hydrodynamic diffusion, and $\chi(\rho)$ is the mobility; they are related via the Einstein relation ${\cal D}(\rho) = f''\,\chi(\rho)$, where $f$ is the free energy density.\\
 As illustration we incorporate the case of the (weakly bulk driven) exclusion process on a ring, which is still diffusive.  There, in the  notation of \eqref{LGEN}, 
 \begin{eqnarray}
\psi^*(\rho,f) &=& \frac 1{2} (f,\chi(\rho)f), \qquad \text{  for  }\;\;
\chi(\rho)= \rho(1-\rho),\label{mob}\\
F(\rho) &=& {\cal E} - \nabla \frac{\delta{\cal F}}{\delta \rho(r)},\; \quad {\cal F}(\rho) = T \int \id r\,[\rho(r)\log \rho(r) + (1-\rho(r))\log(1-\rho(r))]\notag
\end{eqnarray}
where  ${\cal E}$ is the external driving and ${\cal F}(\rho)$ is the free energy for density profile $\rho$ of the reference equilibrium system dominated by the exclusion interaction.\\
Boundary driven diffusion processes have the disadvantage, with respect to the present program, that the fluctuation structure remains quadratic, hence is less rich and remains less transparent for distinguishing the canonical structure in more general nonequilibrium scenarios. \\

 {\bf 4:} Suppose that we consider a typical path $\omega$ where initially $z_i = z(0)$ and finally $z_f =z(t)$ (solution of the zero-cost flow at time $t$ when starting in $z_i$).  The conditional probability of this trajectory can be compared with the time-reversed trajectory $\theta \omega$, and we get from local detailed balance that
\begin{equation}
\text{Prob}[\theta \omega]\,|\,z(t)=z_f] =
\frac{\text{Prob}[\theta \omega\,|\,z(t)=z_f]}{\text{Prob}[\omega\,|\,z(0)=z_i]}
=  e^{-N\int_0^tj(s)\cdot F(z(s))\,\id s}
\end{equation}
where we have used that the trajectory $\omega: z_i \rightarrow z_f$ is pure relaxation under the zero-cost flow.
That shows that an increase of the nonequilibrium entropy (as in $\theta\omega$) is very unlikely, exponentially small in the entropy change of the World (during $\omega$).  That extends the equilibrium case where $\int_0^tj(s)\cdot F(z(s)) = \cal F(z_i) - \cal F(z_f) \geq 0$ (along $\omega$) for the equilibrium free energy density $\cal F$ of the system.

\section{General argument}\label{set}

The idea of connecting the relaxation structure with dynamical large deviations is not original and found in several papers, including \cite{MPR13,ad1,ad2,frank,genericmaes,Kraaij_2020,pons}, while the return to nonequilibrium has not been considered.

\subsection{Setup}
We denote by $z$ a macroscopic state or condition; mathematically, we think of it as an element of a differentiable manifold.  Physically, $z$ may stand for a density profile in a many-particle system, or for a list of concentrations of chemical species in some reactor, {\it etc}.  It evolves in time, with $z(s)$ the macroscopic state at time $s\in (0,t)$.  We assume that the time-evolution is described by a first-order equation of the form
\begin{equation}\label{mee}
\dot z = D\, j_z
\end{equation}
for some current $j_z$,
with $D$ being minus a divergence (like in a continuity equation) or a stochiometric matrix (for chemical reaction networks), or the identity operator (or still some other operator) acting on the current.  The adjoint of $D$ is denoted by $D^{\dagger}$ and satisfies $a \cdot D b = b \cdot D^{\dagger} a$. Note that the scalar product may contain an integration over space.  The question we address in the present paper is to ask what exactly is $j_z$ in \eqref{mee}, the typical current when in state $z$.  Through what structure does it depend on $z$?\\

The approach we take to the last question is that of dynamical fluctuation theory.  We imagine many (in principle) possible trajectories $(z(s),j(s)), s\in (0,t),$ of states and currents, all connected via $\dot z = D\, j(s)$, and we write their probability as
\begin{align}\label{large deviation prob z}
   &\mathbb{P}\Big[ \{z(s), j(s) \}_{0 \leq s \leq t}  \Big] \propto %e^{- N \mathcal{A}\left(\gamma \right)}, \quad \mathcal{A}(\gamma) 
    \exp\left\{ - N\left({\cal V}(z(0)) + \int_{0}^{t} \id s \,\cal L\big(j(s), z(s) \big)\right) \right\} 
\end{align}
as the size or number of components $N\to \infty$. The latter decides the macroscopic regime, much in the spirit of Onsager-Machlup theory, \cite{Onsager1951,Onsager1953}.  
 The Lagrangian $\cal L (j,z)$ of the system is the integrand of the action  
\begin{equation}\label{ax}
\mathcal{A} = \int_{0}^{t} \id s \,\cal L\big(j(s), z(s) \big)
\end{equation}
and the initial state $z(0)$ is sampled from an initial distribution $\exp[-N{\cal V}]$ which can be anything.  Note that there may exist further deterministic constraints $\cal N(j,z)=0$ between current and state, for instance $\dot q=p$ in the case of a Hamiltonian dynamics where the velocity $\dot q$ always follows the momentum $p$ which itself may be stochastic.\\

Probabilities in \eqref{large deviation prob z} arise from a coarse-grained or reduced description. The evolution equation \eqref{mee} is interpreted as generating trajectories that can be characterized as ``typical'' within the set of possible trajectories.  The latter all satisfy $\dot z = Dj$ for some current $j$, but not necessarily equal to $j_z$. We want to characterize the true macroscopic evolution \eqref{mee} as the ``zero-cost flow'' for \eqref{ax}.   We make a couple of natural assumptions:\\
We suppose that $\cal L(j,z)\geq 0$ is strictly convex in $j$. That is the typical situation in (dynamical) large deviation theory, where $\cal L(j,z)$ can be called a dynamical free energy. Obtaining the `equation of motion'  \eqref{mee} is therefore a minimization problem: $\cal L (j, z) = 0$ gives the most likely current $j = j_z$ when in state $z$.\\
Secondly, we assume {\it local detailed balance} in the sense that there exists a flow function $J^H$, so that for all $(z,j)$,
\begin{equation}\label{lbl}
\cal L\left(J^H(z)- j, z \right) - \cal L\left(J^H(z) + j, z \right) = j\cdot F(z)
\end{equation}
for a {\it thermodynamic force}  $F$ (including statistical forces), function of state $z$, with $J^H(z)\cdot F(z) = 0$ (orthogonality). Under well-understood physical conditions of the coupling of the system with the environment, \cite{ldb,time}, the right-hand side in \eqref{lbl} is the (irreversible) entropy flux generated by the force $F$.  The left-hand side gives the change under time-reversal along $J^H$.  That $J^H$ is called the Hamiltonian flow and is zero when the dynamics is purely dissipative.  The introduction of $J^H$ allows to describe an irreversible dissipation around a conservative flow.\\

Let us end this setup with some very brief remark about the mathematics.  The states $z$ belong to some differentiable manifold $ \cal M$, and at each $z\in \cal M$ we find a tangent plane $T_z\cal M$ which contains the  thermodynamic fluxes $j$.  In fact, when $s\mapsto z(s)$ is a smooth curve along $\cal M$, then the time-derivative $\dot z(s) \in T_{z(t)}\cal M$ is a tangent vector.  Dual to the tangent plane $T_z\cal M$ is the cotangent plane $T^*_z\cal M$ and its elements are the driving forces $f$, which can be viewed as linear functionals $j\cdot f$ on tangents $f$.\\
If we have a function $S$ on $\cal M$, we write $\id S$ for its (Gateaux) derivative, which may take various forms of functional derivative.
It is natural to use the language of differential geometry, but we have tried to avoid it

\subsection{Canonical structure}\label{cans}
We repeat the arguments previously used in \cite{sdiv,maes2007entropy,Maes_2008} to establish a canonical structure in the dynamical fluctuation.  It means that the Lagrangian $\cal L$ (as function of the current $j$) can be written as the sum of convex-conjugate functions plus a linear term.  That is crucial (in the next section) to identify the zero-cost flow $j_z$, just as in the case of equilibrium,\cite{genericmaes,Kraaij_2020}.\\

We consider the frenetic part \cite{fren} of the Lagrangian; that is what governs the time-symmetric fluctuations around the Hamiltonian flow,
\[
\frac 1{2}[\cal L(J^H(z)+j,z) + \cal L(J^H(z)-j,z)] = \psi(j,z) + {\cal L}(J^H(z),z) 
\]
which defines the function $\psi$.  Local detailed balance \eqref{lbl} can then be rewritten as
\begin{equation}\label{ps}
\psi(j,z) = \frac 1{2} j\cdot F(z) + \cal L(J^H(z)+j,z) - \cal L(J^H(z),z)
\end{equation}
It appears that $\psi(0,z)=0$ and $\psi(j,z)=\psi(-j,z)$, again from local detailed balance \eqref{lbl}.  Moreover, $\psi$ is convex in $j$ (as inherited from the Lagrangian), and hence $\psi(j,z)\geq 0$.\\
The Legendre transform of $\psi$ is
\begin{equation}\label{sta}
\psi^\star(f,z) = \sup_j\left[j\cdot f - \psi(j,z)\right]
\end{equation}
which we call the {\it dissipation function} (even though it is part of the frenetic contribution to the Lagrangian).  It is convex, symmetric in $\pm f$, vanishes only at $f=0$ and hence is also positive. Furthermore, putting \eqref{ps} into \eqref{sta} we get
 for $f(z)=F(z)/2$, 
\begin{eqnarray}\label{lo}
\cal L(J^H(z),z) &=& \psi^\star(\frac{F(z)}{2},z),\\
\cal L(J^H(z)+j,z) &=& -\frac{1}{2}\,j\cdot F(z) + \psi(j,z) +\psi^\star(\frac{F(z)}{2},z)\nonumber
\end{eqnarray}

\subsection{Relaxation equation}\label{rele}
From \eqref{lo} we conclude that 
\begin{equation}
    \cal L(j,z)=\psi(j-J^H(z),z)+\psi^\star(\frac{F(z)}{2},z)-(j-J^H(z))\cdot \frac{F(z)}{2}\label{LgenGEN}
    \end{equation}
It is (basically) Eq.~1.3 in \cite{patterson2023variational}.\\
Putting $\cal L(j,z)=0$ to find the zero-cost flow $j_z$ thus saturates the Fenchel-Young inequality, 
\[
\psi(j-J^H(z),z)+\psi^\star(\frac{F(z)}{2},z)-(j-J^H(z))\cdot \frac{F(z)}{2} = 0
\]
The solution for $j$
is given by
\begin{equation}\label{fro}
j_z = %\partial \cal H(0,z)= 
J^H(z)+\partial_f\psi^\star(\frac{F(z)}{2},z)
\end{equation}
($\partial_f$ denotes the derivative with respect to the first argument in $\psi^*$), or
\begin{equation}\label{mo}
\dot z=D\,J^H(z) + D\,\partial_f\psi^\star(\frac{F(z)}{2},z)
\end{equation}
which is \eqref{rel}. 

\section{Constructive aspects}\label{const}
The relaxation structure allows us to explore possible scenarios away from specific examples.  In particular, we do not necessarily need to imagine a microscopic or mesoscopic particle system and to first find its Lagrangian governing the macroscopic dynamical fluctuations.  We can directly propose a macroscopic relaxation equation as long as it is compatible with \eqref{mo}, \eqref{rel}.\\

For example, we are perfectly allowed to claim a type of macroscopic overdamped relaxation of a probe with position $z$ that is pumped by a constant force $\cal E$ around a circular tube containing a viscous fluid, by following the structure \eqref{rel}.  Once we specify the thermodynamic force $\cal E$, all depends on the choice of the frenetic part in the Lagrangian.\\
Let us take \eqref{sta} in the form
\begin{equation}\label{cho}
\psi^*(f,z) = \chi_\beta(z,\cal E)\,\{\cosh (f/f_o)-1\}
\end{equation}
for $\chi_\beta(z,\cal E)>0$ (periodic in $z$) and where $f_o$ is a reference force; for example, $f_o^{-1}= L\beta$ where $L$ is the size of the loop and $\beta$ the inverse temperature of the environment.  For thermodynamic force, we take $F(z)= -\partial_z E(z) + \cal E$ with $E$ a periodic energy landscape.  That force $F$ is rotational whenever the constant force $\cal E\neq 0$.  The choice \eqref{cho} implies that
\begin{equation}\label{cir}
\dot z = f_o^{-1}\,\chi_\beta(z, \cal E) \,\sinh ([-\partial_z E(z) + \cal E]/(2f_o))
\end{equation}
which is of the form \eqref{mo},\eqref{rel} for $D$ the unit and $J_z^H=0$. We trust the physical soundness of this (overdamped) dynamical equation \eqref{cir}, generalizing \eqref{mob}, even though no derivation from a specific micro- or mesoscopic model is being considered; local detailed balance is all what is needed. Obviously, \eqref{cir} allows a current over the loop (for $\cal E$ large enough) and, depending on the dependence of $\chi_\beta$ on $\cal E$, may show negative differential mobility.  The dependence on $z$ in $\chi_\beta$ can also be used to select a region in the tube where the probe resides for a longer time, (again) for large enough $\cal E\neq 0$.  For $\cal E=0$, the probe will become stuck in a minimum of $E$. All these aspects can be read from \eqref{cir} without relying on specific microscopic modeling.\\

Obviously, the choice \eqref{cho} is already specific.  As a further constructive and even more phenomenological approach, one can imitate the Landau program for macroscopic order parameters, \cite{goldenfeld1992lectures}, to extend it to nonequilibrium dynamics.    For example, we can start from the convex polynomial
\[
\psi^*(f,z) = a(z)\, f^2 + b(z)\, f^4, \quad a,b\geq 0
\]
and leave the state functions $a, b$ to depend on the thermodynamic force(s) and other parameters as well.  Here, symmetry considerations would enter, much in the spirit of Landau theory, and in the end we can expand around critical points {\it etc.} and construct $\psi^*$ as convex envelope of an analytic but possibly nonconvex polynomial.\\
Note still that the Hohenberg-Halperin classification of (near) critical dynamics deals with time-dependent Ginzburg-Landau theory, which is mostly purely dissipative (gradient flow) and reversible. In that sense, it is a subclass of stochastic gradient flow and of {\tt GENERIC} where white noise got added.  Therefore, formally speaking, our scheme also provides an algorithm to turn this Ginzburg-Landau dynamics into a bulk nonequilibrium dynamics within the larger frame of local detailed balance.  As such, that does not require an {\it a priori} understanding of nonequilibrium criticality, and one replaces in the current the gradient/derivative of the Landau functional by adding the appropriate thermodynamic driving force to study changes in the phase diagram.\\

As a final remark, we emphasize that the discussed dynamics must be distinguished from the (even) more coarse-grained evolution of the reaction coordinates or order parameters.  Even for relaxation to equilibrium and by convexity of the free energy, gradient flow excludes, for example, the presence of (kinetic) energy barriers (at least outside mean-field treatments). Gradient flow captures the possible existence of metastable states (local minima), but not the transitions between them. In other words, the discussed relaxation is not the evolution as usually understood by physical chemistry (transition-path theory) for the evolution of a reaction coordinate in an effective energy profile. The study of the dynamics of reaction or collective coordinates is more related to reaction rate theory, for which a systematic study of the influence of nonequilibrium driving remains largely open, \cite{Hanggi1990,BouchetReygner2016,E_VandenEijnden2006}.

\section{Examples}\label{exam}
To illustrate the notation, we give two examples of return to macroscopic nonequilibrium.\\

\begin{example}[Underdamped motion with a rotational force]
We consider a macroscopic probe of mass $m=1$ with phase space coordinates $z = (q,p)$ in $d\geq 2$ dimensions.  We assume a spatially confining potential $V$ and we add a rotational force $G$.
The underdamped equation of motion with damping coefficient $\gamma > 0$ reads
\begin{equation}\label{ex1}
\dot q= p,\qquad \dot p = -\nabla_q V + G(q) -\gamma \,p
\end{equation}
The stationary macroscopic state is taken to be $z^s = (0,0)$ with $G(0)=0$ and $V(q)$ strictly convex around its minimum at $q=0$.  We are allowed and even encouraged to think that both $V$ and $\gamma$ depend on certain aspects of the force $G$.\\

The thermodynamic force is 
\begin{equation}\label{for1}
F(q,p) =  \begin{bmatrix}
    -\nabla_q V(q)\, & \,\frac{G(q)}{\gamma} - p
  \end{bmatrix}
  \end{equation}
and the dissipation function is  $\psi^*((f_q,f_p),z) = \gamma\,f_p^2$ for force $f=(f_q,f_p)$.\\
Indeed, 
 \eqref{ex1} can be written in the form \eqref{mo} with $D=$ unit 
 operator and 
\begin{eqnarray}
\dot z &=&
\begin{bmatrix}
    \dot q \\ \dot p
  \end{bmatrix} = \begin{bmatrix}
    p \\ -\nabla_q V 
  \end{bmatrix}
+
\begin{bmatrix}
0 \\ G(q) -\gamma\,p
  \end{bmatrix}\nonumber\\
&=& J^H +  \begin{bmatrix}
\partial_1\psi^*(\frac{F}{2}) \\ \partial_2\psi^*(\frac{F}{2}) 
  \end{bmatrix} 
  \end{eqnarray}
for Hamiltonian flow 
\[
J^H(q,p) =
\begin{bmatrix}
    p \\ -\nabla_q V(q) 
  \end{bmatrix},\quad \text{ Hamiltonian } \;H = p^2/2 + V(q)
\]
and $\partial_1\psi^* = 0,\quad\partial_2\psi^*(f_q,f_p) = 2\gamma\,f_p$.\\ 
  We observe that 
\begin{equation}\label{ort}
\nabla_q V \cdot G =0
\end{equation} 
(following the Helmholtz decomposition of the force $G -\nabla_q V $) 
implies the orthogonality $F\cdot J^H = 0$, indicating that the Hamiltonian flow does not produce entropy.\\
We give the canonical form of dynamical large deviations corresponding to a Langevin equation with noise of amplitude $2\gamma/N$ (added to the equation for $\dot p$ in \eqref{ex1}).  We then get the quadratic Lagrangian $\cal L =\cal L(j,z)$ with $j=(\dot q,\dot p), z=(q,p)$,
  \begin{equation}\label{lau}
      {\cal L}= \frac 1{4\gamma}(\dot p + \nabla_qV - G(q) +\gamma p)^2
  \end{equation}
  with constraint $\dot q=p$.  We rewrite \eqref{lau} as
  \begin{equation}\label{lau1}
            \cal L = \frac 1{4\gamma}(\dot p + \nabla_qV)^2 + \frac{\gamma}{4}(\frac{G(q)}{\gamma} - p)^2 - \frac 1{2}(\dot p +\nabla_qV)\cdot(\frac{G(q)}{\gamma} - p)
  \end{equation} 
  Note that $\psi(j,z)= j_p^2/(4\gamma), \psi^*(f,z) = \gamma f_p^2$ and  $j-J^H= (\dot q - p =0,\dot p + \nabla_q V)$, so that the terms in the Lagrangian correspond to
  \begin{eqnarray}
      \frac 1{4\gamma}(\dot p + \nabla_qV)^2 &=& \psi(j-J^H)\\
       \frac{\gamma}{4}(\frac{G(q)}{\gamma} - p)^2 &=&  \psi^*(\frac{F}{2})\\
       \frac 1{2}(\dot p +\nabla_qV)\cdot(\frac{G(q)}{\gamma} - p) &=&  (j_p-J^H_p)\,\frac{F_p(q,p)}{2}
  \end{eqnarray}
  with force $F$ given in \eqref{for1}.  Therefore, the Lagrangian \eqref{lau}--\eqref{lau1}
 indeed equals the canonical form \eqref{LgenGEN}, using the constraint $\dot q=p$.\\
 Note that, in general, there are different Lagrangians that give the same zero-cost flow, but the fluctuations may be different.
\end{example}

\begin{example}[Driven Vlasov--Fokker-Planck equation]\label{vlas}
As variable $z$, we take the time-dependent density $\rho(q,p,t)$ depending on positions $q\in \bbR^d$ and momenta $p\in \bbR^d$ of interacting particles that are treated in a mean-field approximation.  The evolution equation is the driven Vlasov-Fokker-Planck equation and is given by
\begin{equation}\label{vfp}
\dot \rho = -\nabla_q\cdot\rho\frac{p}{m} + \nabla_p\cdot\rho\left(\nabla_qV - G(q) + \nabla_q(\Phi\star\rho)+\gamma\frac{p}{m}\right) + \gamma \,\beta^{-1}\,\Delta_p\rho
\end{equation}
for damping coefficient $\gamma$, mass $m$ and inverse temperature $\beta = (k_BT)^{-1}$.
The convolution is defined as
\[ %\]\begin{equation}
\Phi\star\rho\;(q) = \int_{\bbR^{2d}} \Phi(q-q')\,\rho(q',p')\,\id q'\,\id p'
\] %\end{equation}
and represents the mean-field interaction.   We assume that the rotational force $G$ is orthogonal to the conservative part: $G\cdot \nabla_q W =0$, where, for short, we write $W = V +  \Phi\star\rho$.  Again, $W$ and $\gamma$ may very well depend on $G$.\\
In \eqref{mo} we take
\begin{equation}
D=-\nabla=-\begin{bmatrix}~\nabla_q~&~ \nabla_p~\end{bmatrix}
\end{equation}
representing minus the divergence.
The Hamiltonian flow is 
\[
J^H_\rho = \rho(q,p)
\begin{bmatrix}
    p/m \\ -\nabla_q [V +  \Phi\star\rho](q)
  \end{bmatrix}  %\quad \text{ for } H[\rho] = \int_{\bbR^{2d}}\left(\frac{p^2}{2m} + V(q) + \frac 1{2}(\psi\star\rho)(q)\right)\rho\,\id q\id p
  =  \rho(q,p)\, K\,\nabla \frac{\delta{H}[\rho]}{\delta \rho(r)} 
\]
with antisymmetric $2d\times 2d$ matrix
\begin{equation}\label{mat}
K =\begin{bmatrix} ~~0~& ~1~ \\ -1~& ~0~ \end{bmatrix}
\end{equation}
and energy 
\begin{equation}
H[\rho] = \int_{\bbR^{2d}}\left(\frac{p^2}{2m} + V(q) + (\Phi\star\rho)(q)\right)\,\rho(q,p)\,\id q\id p
\end{equation}
The canonical structure of fluctuations around the Vlasov--Fokker-Planck equation has been established in \cite{Duong2013GENERIC}.
The dissipation function is quadratic $\psi^*(f,\rho) = f\cdot\chi f$ for $2d\times 2d$ matrix
\begin{equation}\label{mate}
\chi=\rho\gamma\begin{bmatrix} ~0~&~0~ \\~0~& ~1~ \end{bmatrix}
\end{equation}
The force vector is
\begin{equation}\label{for}
F(\rho) =  -\nabla\frac{\delta{\cal F}[\rho]}{\delta \rho(r)} + \begin{bmatrix}
    -\nabla_q W \,&\, \frac{G(q)}{\gamma} - p/m
  \end{bmatrix}
  \end{equation}
where the free energy functional is  ${\cal F}[\rho]= H[\rho]- T\, \zeta[\rho]$ with
\begin{equation}
\zeta[\rho] = -k_B \int_{\bbR^{2d}}\rho\log\rho\,\id q\id p
\end{equation}
The evolution \eqref{vfp} can now be written as in \eqref{mo} and the reader easily checks the orthogonality $F \cdot J^H=0$.
\end{example}

\section{Conclusion}

We have shown in what way a driven macroscopic system returns to its stationary nonequilibrium condition, {\it e.g.}, in the presence of rotational thermodynamic forces.  The relaxation equation is not variational and there appears a structure that generalizes {\tt GENERIC}, the mold for relaxation to equilibrium.  That structure not only gives common ground to research programs on phenomena of nonequilibrium relaxation (glassy behavior,\cite{maliet2025bacterialglasstransition}, metastability, \cite{Hurtado_2003}, and localization phenomena, \cite{MORI20083684}).  It extends the Onsager program for connecting relaxation and fluctuation behavior.  Indeed, the analysis is based on the recognition of a canonical structure in the functional or Lagrangian governing macroscopic dynamical fluctuations. As a consequence, on this macroscopic and dynamical level of description, a firm relation is established between the relaxation and the fluctuation structure.  The frenetic contribution in the Lagrangian (time-symmetric part) decides the relaxational behavior, and {\it vice versa}.\\

\vspace{0.5cm}

\noindent{\bf Acknowledgment:}  We are grateful to Mark Peletier and to Andr\'e Schlichting for comments and suggestions.

%%%%%%%%%%%%%%%%%%%
\bibliographystyle{unsrt}  
\bibliography{allpapers}

@article{M,
title={Macroscopic fluctuation theory},
volume={87},
ISSN={1539-0756},
url={http://dx.doi.org/10.1103/RevModPhys.87.593},
DOI={10.1103/revmodphys.87.593},
number={2},
journal={Reviews of Modern Physics},
publisher={American Physical Society (APS)},
author={Bertini, L. and De Sole, A. and Gabrielli, D. and Jona-Lasinio, G. and Landim, C.},
year={2015}, pages={593–636} }

@article{F,
title = {{Fluctuations in Stationary Nonequilibrium States of Irreversible Processes}},
author = {Bertini, L. and De Sole, A. and Gabrielli, D. and Jona-Lasinio, G. and Landim, C.},
journal = {Physical Review Letters},
volume = {87},
issue = {4},
pages = {040601},
numpages = {4},
year = {2001},
publisher = {American Physical Society},
doi = {10.1103/PhysRevLett.87.040601},
url = {https://link.aps.org/doi/10.1103/PhysRevLett.87.040601}
}

@article{T,
title = {Macroscopic fluctuation theory for stationary non equilibrium states},
volume={107},
ISSN={0022-4715},
url={http://dx.doi.org/10.1023/A:1014525911391},
DOI={10.1023/a:1014525911391},
number={3/4},
journal={Journal of Statistical Physics},
publisher={Springer Science and Business Media LLC},
author={Bertini, L. and De Sole, A. and Gabrielli, D. and Jona-Lasinio, G. and Landim, C.},
year={2002},
pages={635–675} }

@article{Kirkwood1946,
  author    = {Kirkwood, J.G.},
  title     = {The Statistical Mechanical Theory of Transport Processes. I. General Theory},
  journal   = {The Journal of Chemical Physics},
  year      = {1946},
  volume    = {14},
  number    = {3},
  pages     = {180--201},
  doi       = {10.1063/1.1724117}
}

@Article{ldb,
	title={{Local detailed balance}},
	author={C. Maes},
	journal={SciPost Phys. Lect. Notes},
	pages={32},
	year={2021},
	publisher={SciPost},
	doi={10.21468/SciPostPhysLectNotes.32},
	url={https://scipost.org/10.21468/SciPostPhysLectNotes.32},
}

@article{aaron,
url = {https://doi.org/10.1515/jnet-2024-0054},
title = {Entropy as {Noether} charge for quasistatic gradient flow},
title = {},
author = {A. Beyen and C. Maes},
journal = {Journal of Non-Equilibrium Thermodynamics},
doi = {doi:10.1515/jnet-2024-0054},
year = {2025},
lastchecked = {2025-03-13}
}

@book{Spohn1991,
  author    = {H. Spohn},
  title     = {Large Scale Dynamics of Interacting Particles},
  year      = {1991},
  publisher = {Springer},
  series    = {Theoretical and Mathematical Physics},
  doi       = {10.1007/978-3-642-84371-6},
  url       = {https://link.springer.com/book/10.1007/978-3-642-84371-6}
}

@article{Hanggi1990,
  author       = {P. H\"anggi and P. Talkner and M. Borkovec},
  title        = {Reaction-rate theory: fifty years after {Kramers}},
  journal      = {Reviews of Modern Physics},
  year         = {1990},
  volume       = {62},
  number       = {2},
  pages        = {251--341},
  doi          = {10.1103/RevModPhys.62.251},
}

@article{BouchetReygner2016,
  author       = {F. Bouchet and J. Reygner},
  title        = {Generalisation of the {Eyring–Kramers} transition rate formula to irreversible diffusion processes},
  journal      = {Annales Henri Poincaré},
  year         = {2016},
  volume       = {17},
  number       = {12},
  pages        = {3499--3532},
  doi          = {10.1007/s00023-016-0507-4},
}

@article{E_VandenEijnden2006,
  author       = {E. Weinan and E. Vanden‑Eijnden},
  title        = {Towards a theory of transition paths},
  journal      = {Journal of Statistical Physics},
  year         = {2006},
  volume       = {123},
  number       = {3},
  pages        = {503--523},
  doi          = {10.1007/s10955-005-9003-9},
}

@article{Bertini_2013,
	doi = {10.1103/physrevlett.110.020601}, 
	url = {https://doi.org/10.1103%2Fphysrevlett.110.020601}, 
	year = {2013},
	publisher = {American Physical Society ({APS}},  
	volume = {110},
	number = {2},
	author = {L.  Bertini and D.  Gabrielli and G.  Jona-Lasinio and C.  Landim},
	title = {Clausius Inequality and Optimality of Quasistatic Transformations for Nonequilibrium Stationary States},
	journal = {Phys.  Rev. ~Lett.},
}

@article{fren,
	doi = {10.1016/j.physrep.2020.01.002},
  	url = {https://doi.org/10.1016%2Fj.physrep.2020.01.002},
  	year = {2020},
  	publisher = {Elsevier {BV}},
  volume = {850},
 	pages = {1--33},
 	author = {C. Maes},
 	title = {Frenesy: Time-symmetric dynamical activity in nonequilibria},
 	journal = {Physics Reports},
}

@article{prig,
  author = {P. Glansdorff and G. Nicolis and I. Prigogine},
  title = {The Thermodynamic Stability Theory of Non-Equilibrium States},
  journal = {Proc. Nat. Acad. Sci. USA},
  volume = {71},
  pages = {197--199},
  year = {1974},
}

@article{GP,
  author = {C. Maes and K. Neto\v{c}n\'{y}},
  title = {Revisiting the {Glansdorff-Prigogine} criterion for stability within irreversible thermodynamics},
  journal = {J. Stat. Phys.},
  volume = {159},
  pages = {1286--1299},
  year = {2015},
}

@article{Onsager1951,
  author       = {L. Onsager and S. Machlup},
  title        = {Fluctuations and Irreversible Processes. {I}},
  journal      = {Physical Review},
  volume       = {91},
  number       = {6},
  pages        = {1505--1512},
  year         = {1953},
  doi          = {10.1103/PhysRev.91.1505}
}

@article{Onsager1953,
  author       = {S. Machlup and L. Onsager},
  title        = {Fluctuations and Irreversible Processes. {II}. Systems with Kinetic Energy},
  journal      = {Physical Review},
  volume       = {91},
  number       = {6},
  pages        = {1512--1515},
  year         = {1953},
  doi          = {10.1103/PhysRev.91.1512}
}

@book{goldenfeld1992lectures,
  title={Lectures on Phase Transitions and the Renormalization Group},
  author={Goldenfeld, N.},
  year={1992},
  publisher={Addison-Wesley},
  series={Frontiers in Physics},
  address={Reading, Massachusetts},
  isbn={978-0201554090}
}

@misc{maliet2025bacterialglasstransition,
      title={Bacterial Glass Transition}, 
      author={M. Maliet and N. Fix-Boulier and L. Berthier and M. Deforet},
      year={2025},
      eprint={2504.04205},
      archivePrefix={arXiv},
      primaryClass={cond-mat.soft},
      url={https://arxiv.org/abs/2504.04205}, 
}

@inproceedings{Hurtado_2003,
   title={Metastability and Avalanches in a Nonequilibrium Ferromagnetic System},
   volume={661},
   ISSN={0094-243X},
   url={http://dx.doi.org/10.1063/1.1571303},
   DOI={10.1063/1.1571303},
   booktitle={AIP Conference Proceedings},
   publisher={AIP},
   author={P.I. Hurtado and J. Marro and P.L Garrido},
   year={2003},
   pages={147–152} 
}

@article{MORI20083684,
title = {Wave-Pinning and Cell Polarity from a Bistable Reaction-Diffusion System},
journal = {Biophysical Journal},
volume = {94},
number = {9},
pages = {3684-3697},
year = {2008},
issn = {0006-3495},
doi = {https://doi.org/10.1529/biophysj.107.120824},
url = {https://www.sciencedirect.com/science/article/pii/S0006349508704442},
author = {Y. Mori and A. Jilkine and L. Edelstein-Keshet}
}

@article{maes1999fluctuation,
  author    = {Maes, C.},
  title     = {The Fluctuation Theorem as a {Gibbs} Property},
  journal   = {Journal of Statistical Physics},
  volume    = {95},
  number    = {1-2},
  pages     = {367--392},
  year      = {1999},
  publisher = {Springer},
  doi       = {10.1023/A:1004541830998}
}

@article{maes2000entropy,
  author    = {Maes, C. and Redig, F. and Van Moffaert, A.},
  title     = {On the definition of entropy production, via examples},
  journal   = {Journal of Mathematical Physics},
  volume    = {41},
  number    = {3},
  pages     = {1528--1554},
  year      = {2000},
  publisher = {AIP Publishing},
  doi       = {10.1063/1.533195}
}

@BOOK{gasp,
       author = {{Gaspard}, P.},
        title = "{The Statistical Mechanics of Irreversible Phenomena}",
         year = 2022,
          doi = {10.1017/9781108563055},
publisher = {Cambridge University Press},
  year = {2022},
}

@article{time,
  author = {C. Maes and K. Neto\v{c}n\'y},
  title = {Time-reversal and Entropy},
  journal = {J. Stat. Phys.},
  volume = {110},
  pages = {269},
  year = {2003},
}

@article{sdiv,
   title={Steady state statistics of driven diffusions},
   volume={387},
   ISSN={0378-4371},
   url={http://dx.doi.org/10.1016/j.physa.2008.01.097},
   DOI={10.1016/j.physa.2008.01.097},
   number={12},
   journal={Physica A: Statistical Mechanics and its Applications},
   publisher={Elsevier BV},
   author={C. Maes and K. Neto\v{c}n\'{y} and  B. Wynants},
   year={2008},
pages={2675-2689} 
}

@book{fw,
  title={Random Perturbations of Dynamical Systems},
  author={M. I. Freidlin and A. D. Wentzell},
  year={1998},
  publisher={Springer-Verlag},
  series={Grundlehren der Mathematischen Wissenschaften},
  number={260},
  address={New York},
}

@book{dGM,
    author    = {De Groot, S. R. and Mazur, P.},
    title     = {Non-Equilibrium Thermodynamics},
    year      = {2013},
    publisher = {Courier Corporation},
    address   = {Dover Books on Physics},
    edition   = {reprint},
    isbn      = {9780486139477}
}

@article{BasuMaesNetocny2015,
  author  = {Basu, U. and Maes, C. and Neto{\v{c}}n{\'y}, K.},
  title   = {Statistical forces from close-to-equilibrium media},
  journal = {New Journal of Physics},
  volume  = {17},
  pages   = {115006},
  year    = {2015},
  doi     = {10.1088/1367-2630/17/11/115006}
}

@article{PhysRevResearch.4.033208,
  title = {Hessian geometry of nonequilibrium chemical reaction networks and entropy production decompositions},
  author = {Kobayashi, T.J. and Loutchko, D. and Kamimura, A. and Sughiyama, Y.},
  journal = {Phys. Rev. Res.},
  volume = {4},
  issue = {3},
  pages = {033208},
  numpages = {17},
  year = {2022},
  month = {Sep},
  publisher = {American Physical Society},
  doi = {10.1103/PhysRevResearch.4.033208},
  url = {https://link.aps.org/doi/10.1103/PhysRevResearch.4.033208}
}

@book{DemboZeitouni1998,
  author    = {Dembo, A. and Zeitouni, O.},
  title     = {{Large Deviations Techniques and Applications}},
  series    = {{Stochastic Modelling and Applied Probability}},
  volume    = {38},
  publisher = {Springer},
  address   = {New York},
  year      = {1998},
  edition   = {2},
  doi       = {10.1007/978-1-4612-5320-4}
}

@misc{BodineauLebowitzMouhotVillani2013,
  author       = {Bodineau, T. and Lebowitz, J.L. and Mouhot, C. and Villani, C.},
  title        = {Lyapunov functionals for boundary-driven nonlinear drift-diffusions},
  year         = {2013},
  eprint       = {1305.7405},
  archivePrefix= {arXiv},
  primaryClass = {math.AP},
  url          = {https://arxiv.org/abs/1305.7405}
}

@article{DeRoeckMaesNetocny2006,
  author  = {De Roeck, W. and Maes, C. and Neto{\v{c}}n{\'y}, K.},
  title   = {H-theorems from macroscopic autonomous equations},
  journal = {Journal of Statistical Physics},
  volume  = {123},
  number  = {3},
  pages   = {571--584},
  year    = {2006},
  doi     = {10.1007/s10955-006-9056-8}
}

@article{Maes_2008,
   title={Canonical structure of dynamical fluctuations in mesoscopic nonequilibrium steady states},
   volume={82},
   ISSN={1286-4854},
   url={http://dx.doi.org/10.1209/0295-5075/82/30003},
   DOI={10.1209/0295-5075/82/30003},
   number={3},
   journal={EPL (Europhysics Letters)},
   publisher={IOP Publishing},
   author={C. Maes and K.  Netočný},
   year={2008},
   month=apr, pages={30003} }

@misc{maes2007entropy,
      title={On and beyond entropy production: the case of {Markov} jump processes}, 
      author={C. Maes and K. Netočný and B. Wynants},
      year={2007},
      eprint={0709.4327},
      archivePrefix={arXiv},
      primaryClass={cond-mat.stat-mech},
      url={https://arxiv.org/abs/0709.4327}, 
}

@article{Maes2008Steady,
   title={Steady state statistics of driven diffusions},
   volume={387},
   ISSN={0378-4371},
   url={http://dx.doi.org/10.1016/j.physa.2008.01.097},
   DOI={10.1016/j.physa.2008.01.097},
   number={12},
   journal={Physica A: Statistical Mechanics and its Applications},
   publisher={Elsevier BV},
   author={C. Maes and K. Netočný and B. Wynants},
   year={2008},
   month=may, pages={2675–2689} }

@article{genericmaes,
   title={{Deriving GENERIC from a Generalized Fluctuation Symmetry}},
   volume={170},
   ISSN={1572-9613},
   url={http://dx.doi.org/10.1007/s10955-017-1941-5},
   DOI={10.1007/s10955-017-1941-5},
   number={3},
   journal={Journal of Statistical Physics},
   publisher={Springer Science and Business Media LLC},
   author={Kraaij, R. and Lazarescu, A. and Maes, C. and Peletier, M.},
   year={2017},
   pages={492–508} }

@article{Maes2026WhatIsNonequilibrium,
  title        = {What is nonequilibrium?},
  author       = {Maes, C.},
  journal      = {arXiv preprint},
  eprint       = {2601.16716},
  archivePrefix= {arXiv},
  primaryClass = {cond-mat.stat-mech},
  year         = {2026},
  version      = {v2},
  url          = {https://arxiv.org/abs/2601.16716v2},
  note         = {Lecture notes on nonequilibrium statistical mechanics},
}

@article{complexfluids1,
  title = {{Dynamics and thermodynamics of complex fluids.  I. Development of a general formalism}},
  author = {Grmela, M. and \"Ottinger, H. C.},
  journal = {Phys. Rev. E},
  volume = {56},
  issue = {6},
  pages = {6620--6632},
  numpages = {0},
  year = {1997},
  month = {Dec},
  publisher = {American Physical Society},
  doi = {10.1103/PhysRevE.56.6620},
  url = {https://link.aps.org/doi/10.1103/PhysRevE.56.6620}
}

@phdthesis{Hoeksema2023,
  author       = {Hoeksema, J.},
  title        = {Mean-field limits and beyond: Large deviations for singular interacting diffusions and variational convergence for population dynamics},
  school       = {Eindhoven University of Technology},
  year         = {2023},
  month        = {February},
  isbn         = {978-90-386-5654-0},
  url          = {https://pure.tue.nl/ws/portalfiles/portal/261388064/20230203_Hoeksema_hf.pdf},
  note         = {PhD thesis, Technische Universiteit Eindhoven, defended 3 February 2023}
}

@article{Kaiser2018,
  author    = {Kaiser, M. and Jack, R. L. and Zimmer, J.},
  title     = {Canonical Structure and Orthogonality of Forces and Currents in Irreversible Markov Chains},
  journal   = {Journal of Statistical Physics},
  year      = {2018},
  volume    = {170},
  number    = {6},
  pages     = {1019--1050},
  doi       = {10.1007/s10955-018-1986-0},
  url       = {https://doi.org/10.1007/s10955-018-1986-0},
  issn      = {1572-9613}
}

@article{complexfluids2,
  title = {{Dynamics and thermodynamics of complex fluids.  II. Illustrations of a general formalism}},
  author = {\"Ottinger, H. C. and Grmela, M.},
  journal = {Phys. Rev. E},
  volume = {56},
  issue = {6},
  pages = {6633--6655},
  numpages = {0},
  year = {1997},
  month = {Dec},
  publisher = {American Physical Society},
  doi = {10.1103/PhysRevE.56.6633},
  url = {https://link.aps.org/doi/10.1103/PhysRevE.56.6633}
}

@article{Kraaij_2020,
   title={{Fluctuation symmetry leads to GENERIC equations with non-quadratic dissipation}},
   volume={130},
   ISSN={0304-4149},
   url={http://dx.doi.org/10.1016/j.spa.2019.02.001},
   DOI={10.1016/j.spa.2019.02.001},
   number={1},
   journal={Stochastic Processes and their Applications},
   publisher={Elsevier BV},
   author={R. C. Kraaij and  A. Lazarescu and C. Maes and M. Peletier},
   year={2020},
pages={139–170} }

@article{pons,
   title={{A Generalization of Onsager’s Reciprocity Relations to Gradient Flows with Nonlinear Mobility}},
   volume={41},
   ISSN={1437-4358},
   url={http://dx.doi.org/10.1515/jnet-2015-0073},
   DOI={10.1515/jnet-2015-0073},
   number={2},
   journal={Journal of Non-Equilibrium Thermodynamics},
   publisher={Walter de Gruyter GmbH},
   author={Mielke, A. and Renger, D. R. M. and Peletier, M. A.},
   year={2016},
   month=apr, pages={141–149} }

@article{O,
  title = {{Reciprocal Relations in Irreversible Processes. I.}},
  author = {Onsager, L.},
  journal = {Phys. Rev.},
  volume = {37},
  number = {4},
  pages = {405--426},
  numpages = {0},
  year = {1931},
  month = {Feb},
  publisher = {American Physical Society},
  doi = {10.1103/PhysRev.37.405},
  url = {https://link.aps.org/doi/10.1103/PhysRev.37.405}
}

@article{MPR13,
   title={{On the Relation between Gradient Flows and the Large-Deviation Principle, with Applications to Markov Chains and Diffusion}},
   volume={41},
   ISSN={1572-929X},
   url={http://dx.doi.org/10.1007/s11118-014-9418-5},
   DOI={10.1007/s11118-014-9418-5},
   number={4},
   journal={Potential Analysis},
   publisher={Springer Science and Business Media LLC},
   author={Mielke, A. and Peletier, M. A. and Renger, D. R. M.},
   year={2014},
   month=jun, pages={1293–1327} }

@article{patterson2023variational,
author = {R. I. A. Patterson and D. R. M. Renger and U. Sharma},
year = {2024},
month = {01},
title = {{Variational structures beyond gradient flows: a macroscopic fluctuation-theory perspective}},
volume = {191},
journal = {Journal of Statistical Physics},
doi = {10.1007/s10955-024-03233-8}
}

@article{Duong2013GENERIC,
  author       = {M.H. Duong and M.A. Peletier and J. Zimmer},
  title        = {{GENERIC} formalism of a {Vlasov--Fokker--Planck} equation and connection to large--deviation principles},
  journal      = {Nonlinearity},
  volume       = {26},
  pages        = {2951--2971},
  year         = {2013},
  doi          = {10.1088/0951-7715/26/11/2951}
}

@article{ad1,
   title={{ From a large-deviations principle to the Wasserstein
gradient flow: A new micro-macro passage}},
   volume={307},
   journal= {Commun. Math. Phys.},
   author={Adams, S. and Dirr, N. and Peletier, M.A. and Zimmer, J.},
year={2011},
    pages={791–815} }

@article{ad2,
author={Adams, S. and Dirr, N. and Peletier, M.A. and Zimmer, J.},
year={2013},
    pages={20120341},
   title={{ Large deviations and gradient flows}},
   volume={371},
   journal= {Philos. Trans. Royal
Soc. A.},
    }

@article{frank,
   title={{Large deviations in stochastic heat-conduction processes provide a
gradient-flow structure for heat conduction}},
   volume={55(9)},
   journal= { J. Math. Phys.},
   author={Peletier, M.A. and Redig, F. and Vafayi, K.},
year={2014},
    pages={20120341} }
\onecolumngrid

\end{document}